\documentclass[prb,aps,twocolumn,amsmath,amssymb,floatfix,showpacs,
superscriptaddress]{revtex4}

\usepackage[dvips]{graphics}
\usepackage{color}
\definecolor{dred}{rgb}{0,0,0.6}

\begin{document}

\title{\textcolor{dred}{Multiple mobility edges in a 1D Aubry chain with 
Hubbard interaction in presence of electric field: Controlled electron 
transport}}

\author{Srilekha Saha Saha}

\affiliation{Condensed Matter Physics Division, Saha Institute of Nuclear
Physics, Sector-I, Block-AF, Bidhannagar, Kolkata-700 064, India}

\author{Santanu K. Maiti}

\email{santanu.maiti@isical.ac.in}

\affiliation{Physics and Applied Mathematics Unit, Indian Statistical
Institute, 203 Barrackpore Trunk Road, Kolkata-700 108, India}

\author{S. N. Karmakar}

\affiliation{Condensed Matter Physics Division, Saha Institute of Nuclear
Physics, Sector-I, Block-AF, Bidhannagar, Kolkata-700 064, India}

\begin{abstract}

Electronic behavior of a 1D Aubry chain with Hubbard interaction is 
critically analyzed in presence of electric field. Multiple energy bands 
are generated as a result of Hubbard correlation and Aubry potential,
and, within these bands localized states are developed under the 
application of electric field. Within a tight-binding framework we 
compute electronic transmission probability and average density of 
states using Green's function approach where the interaction parameter
is treated under Hartree-Fock mean field scheme. From our analysis we
find that selective transmission can be obtained by tuning injecting 
electron energy, and thus, the present model can be utilized as a 
controlled switching device. 

\end{abstract}


\maketitle

\section{Introduction}

Electron transport in low-dimensional system has created a lot of interest 
among researchers due to its immense applicability in the field of 
nanoscience. Transport in low-dimensional systems led to interesting 
quantum effects. In one-dimension ($1$D) in presence of random disorder 
all the eigenstates are exponentially localized irrespective of however weak 
is the strength of disorder, this is the well-known phenomenon of Anderson 
localization~\cite{ander}. Based on this fact it is a common belief that no 
mobility edge, energy eigenvalues separating localized states from the 
extended states, can exist in 1D. In addition to 
Anderson localization there exists another kind of localization which 
is Wannier Stark localization~\cite{wann} that occurs due to application 
of bias voltage. It has also drawn much attention like the case of Anderson 
localization. Here localization is obtained even in absence of disorder 
and only due to the resulting electric field. Many theoretical~\cite{wt1,
wt2,wt3,wt4,wt5} and experimental~\cite{wae} analysis are available on 
Stark localization just like Anderson localization. Even in this case 
mobility edge could not be detected. 

However, it has been pointed out that in correlated disordered systems 
all eigenstates are not localized~\cite{angel}, rather some states are 
of extended in nature as well. In a work Dunlap {\em et al.} considered 
a random dimer model~\cite{dunlap} and showed that the system supports 
extended eigenstates at certain discrete eigenvalues. Similarly, a number 
of works have also appeared in the literature~\cite{l1,l2,ssa2,sn1} 
to establish the 
presence of delocalized states along with the localized ones thereby 
exhibiting metal to insulator transition. With all these special classes
of lattice models, Aubry-Andre (AA) model~\cite{abam} always gives a
a classic signature in transport phenomena. The on-site potential in the 
AA 1D chain has the form of a cosine function~\cite{mag,sharma,eco}:
\begin{equation}
\epsilon_n = \lambda \cos (Qna)
\label{aaeq}
\end{equation}
where $\lambda$ is the modulation amplitude, $Q$ is an irrational multiple 
of $\pi$ and $a$ is the lattice spacing. It is a quasiperiodic lattice 
something intermediate between periodic and random disordered systems. 
The parameter $\lambda$ has an important role on the localization behavior 
of the eigenstates. Using Thouless formula~\cite{thf} Aubry and Andre 
demonstrated that this model exhibits energy independent metal insulator 
transition in the parameter space of the Hamiltonian at $\lambda=2t$, 
where $t$ represents the nearest-neighbor hopping integral. For 
$\lambda < 2t$ all eigenstates are extended and in case of 
$\lambda > 2t$ all are localized, the equality relation is the point of 
duality with exotic critical eigenstates which are neither extended nor 
localized. This interesting feature of the AA model aroused immense 
\begin{figure}[ht]
{\centering \resizebox*{7.5cm}{2cm}{\includegraphics{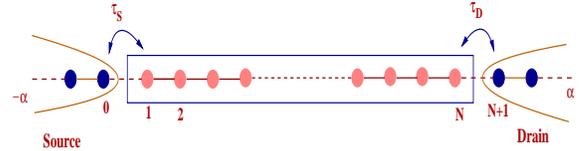}}\par}
\caption{(Color online). Schematic diagram of a $1$D tight-binding chain 
coupled to two $1$D semi-infinite electrodes, viz, source and drain.}
\label{chain}
\end{figure}
curiosity in the minds of researchers to advance in this field. Large 
number of articles dealt with this model both for $1D$ chain as well as 
in case of ladder~\cite{aubry79,ssa1}. However to the best of 
our knowledge, 
the study of the phenomenon of Stark localization in an AA chain in presence 
of Hubbard interaction is absent in literature. In the present manuscript we 
have elaborately studied the phenomena of localization in a 1D AA chain in 
absence and presence of electron-electron (e-e) interaction. First we 
analyze the effect of applied bias voltage on the localization behavior 
of AA chain in absence of any Hubbard interaction and finally we approach 
to the case in presence of interaction. Thus we see how an interesting 
interplay of localization as well as mobility edge phenomena occurring.

Rest of the article is arranged as follows. In Section II we present the 
model and theory based on which the results have been derived and 
discussed in Section III. Lastly we conclude in Section IV.  
    
\section{Model and Theory}

Figure~\ref{chain} depicts a one-dimensional AA chain coupled to two 
semi-infinite leads. The chain comprising $N$ atomic sites is subjected 
to an incommensurate Aubry potential. We describe the model embracing the 
tight-binding formalism and Hamiltonian for the entire system can be 
expressed as,
\begin{equation}
H=H_S + H_{chain} + H_{tun} + H_{D}
\label{eq1}
\end{equation} 
where the different sub-Hamiltonians are described as follows. The 
Hamiltonians
\begin{figure}[ht]
{\centering \resizebox*{6.5cm}{4cm}{\includegraphics{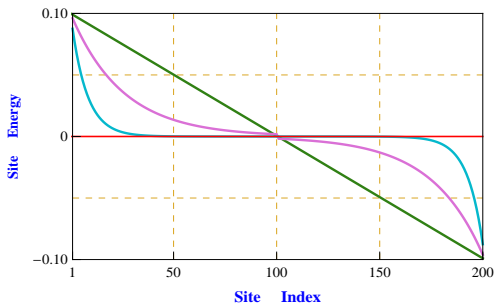}}\par}
\caption{(Color online). Variation of voltage dependent on-site potentials 
in a $1$D chain with $200$ atomic sites when the bias voltage is fixed at
$0.2\,$V. Three different electrostatic potential profiles, one linear and 
two non-linear, are taken into account those are represented by three 
different colored curves.}
\label{efl}
\end{figure}
for the semi-infinite source and drain electrodes are $H_{S(D)}$ and 
they can be written explicitly as,
\begin{equation}
H_{S(D)}=\sum \limits_p \epsilon_0 d_p^{\dag} d_p + 
\sum\limits_p t_0 [d_{p+1}^{\dag} d_p + h.c.]
\label{eq2}
\end{equation}
where $\epsilon_0$ and $t_0$ correspond to the on-site energy
and nearest-neighbor hopping integral, respectively, in the electrodes. 
Creation and annihilation operators of electron inside the electrodes 
in the $n$th Wannier state are respectively denoted by $d_n^{\dag}$ and 
$d_n$.

The second term in Eq.~\ref{eq1} describes the Hamiltonian for the 1D AA 
chain. In absence of electron-electron interaction, the AA chain has on-site 
energy of the form $\epsilon_i=\lambda \cos(Qia)$ where, $a$ represent the 
lattice constant, $Q$ is an irrational multiple of $\pi$ and $i$ corresponds 
to positions of atomic sites. Therefore Hamiltonian for a non-interacting 
AA chain is of the form
\begin{equation}
H_{chain}=\sum \limits_i \epsilon_i c_i^{\dag} c_i + 
\sum\limits_i t [c_{i+1}^{\dag} c_i + h.c.]
\label{eq4}
\end{equation}
$t$ being nearest-neighbor hopping integral and $c_i^{\dag}$ ($c_i$ ) denotes
the creation (annihilation) operator in the AA chain. When all site
energies are identical, the AA chain maps to a perfect ordered chain and
in that case we set $\epsilon_i=0$ $\forall \;i$, without loss of generality. 
As we apply bias voltage $V$ across the chain, it results an electric field 
to develop across it and the on-site energy gets modified to 
$\epsilon_i^{\prime}=\epsilon_i + \epsilon_i(V)$, where $\epsilon_i(V)$ is 
the voltage dependent part of on-site energy.
Now, the question naturally comes how on-site energy 
depends on the voltage. Several attempts~\cite{pr1,pr2,pr3,pr4,pr5,pr6} 
have been done along this line to find potential profile along a linear 
conductor coupled to source and drain electrodes. Mostly {\em ab initio} 
or semiempirical models are used, but few groups have also done it by 
solving Poisson's equations. From these studies it has been suggested 
that depending on electron screening, potential profile along a conductor
sandwiched between source and drain can be of linear type or non-linear
one. For very large screening length linear voltage drop is expected, 
while entire voltage drops at the interfaces between chain and electrode
for very short screening length~\cite{pr5}. Considering symmetric potential
drop at the two interfaces we can write voltage dependent on-site energy 
for linear potential profile as~\cite{pr7} $\epsilon_i(V)$=$V/2 - iV/(N+1)$, 
where 
$V$ is the applied bias voltage. While, for non-linear potential profile
there is no such specific functional form. Keeping in mind the effect of
electron screening, in our model calculations we choose two different 
functional forms for two non-linear curves as shown in Fig.~\ref{efl}.
For pink curve we set $V/2\; \mbox{Exp} [-i V/5]$, while for blue curve 
it gets the form $V/2\; \mbox{Exp} [-i V/1.5]$. Running the variable
$i$ (viz, site index) from $1$ to $N/2$ we generate the datas for fixed
voltage $V$ and then for $N/2+1$ to $N$ values of $i$ we take the negative
of these values in reverse order to make the profile symmetric. One can
also take other functional forms to generate these identical curves and 
the entire physics will remain same as electron transport depends only on
the values of voltage dependent on-site energies, not on the actual 
functional form. Here two non-linear profiles correspond to two different
electron screening as suggested by earlier studies~\cite{pr5,pr6}.
Here it is important to note that, the variation of electrostatic potential 
profile depends on the material itself. But for our model calculations we 
consider these three different profiles, and, we believe that with our 
results general features of electric field on electron transmission across 
a junction can be clearly analyzed. 

The third term represents coupling Hamiltonian due to the coupling of 
the AA chain and side-attached leads, and it reads as
\begin{equation}
H_{tunn}=\tau_{S} [c_1^\dag d_0 + h.c.] + \tau_{D} 
[c_N^\dag d_{N+1} + h.c.]
\label{eq3} 
\end{equation}
where $\tau_{S}$ and $\tau_{D}$ give the strengths by which the system 
is coupled to source and drain, respectively.

Now if we incorporate on-site Coulomb interaction in the AA chain through
Hubbard term along with the effect of bias voltage, the Hamiltonian takes
the form,
\begin{eqnarray}
H_{chain}& = & \sum \limits_{i,\sigma} \epsilon_{i,\sigma}^{\prime} 
c_{i,\sigma}^\dag 
c_{i,\sigma} + \sum \limits_{\langle ij \rangle, \sigma} 
t [c_{i,\sigma}^\dag
c_{j,\sigma} + c_{j,\sigma}^\dag c_{i,\sigma}] \nonumber \\
& + & \sum \limits_i U c_{i \uparrow}^\dag c_{i \uparrow} 
c_{i \downarrow}^\dag c_{i \downarrow}  
\label{eq5}
\end{eqnarray}
where $U$ is the Coulomb interaction strength.

To study electronic behavior of such an interacting system we use 
Hartree-Fock mean field~\cite{kato,kam,new1} theory. In this approach, 
Eq.~\ref{eq5} can be written as 
\begin{eqnarray}
H_{chain} &=&\sum_i \epsilon_{i\uparrow}^{\prime\prime} 
n_{i\uparrow} + 
\sum_{\langle ij \rangle} t \left[c_{i\uparrow}^{\dagger} c_{j\uparrow} + 
c_{j\uparrow}^{\dagger} c_{i\uparrow}\right] \nonumber \\
& + & \sum_i \epsilon_{i\downarrow}^{\prime\prime} n_{i\downarrow} 
+ \sum_{\langle ij \rangle} t \left[c_{i\downarrow}^{\dagger} c_{j\downarrow}
+ c_{j\downarrow}^{\dagger} c_{i\downarrow}\right] \nonumber \\
& - & \sum_i U_i \langle n_{i\uparrow} \rangle \langle n_{i\downarrow} 
\rangle \nonumber \\
&=& H_{c,\uparrow} + H_{c,\downarrow} - 
\sum_i U_i \langle n_{i\uparrow} \rangle \langle n_{i\downarrow} \rangle
\label{eq6} 
\end{eqnarray}   
where, $H_{c,\uparrow}$ and $H_{c,\downarrow}$ are Hamiltonians for the up 
and down spin electrons, respectively, similar to Eq.~\ref{eq4} with 
modified on-site energies as $\epsilon_{i,\uparrow}^{\prime\prime}=\lambda 
\cos(Qia) + \epsilon_i(V) + U \langle n_{i,\downarrow} \rangle$ and 
$\epsilon_{i,\downarrow}^{\prime\prime} =\lambda \cos(Qia) + \epsilon_i(V) 
+ U \langle n_{i,\uparrow} \rangle$, $n_{i,\sigma}=c_{i,\sigma}^\dag 
c_{i,\sigma}$ being the number operator. 
The last term in the above equation (Eq.~\ref{eq6}) represents a shift in 
the total energy
and it depends on the average number of up and down spin electrons.
Once we get the decoupled Hamiltonians for up and down spin electrons, we
find the eigenvalues self-consistently considering some initial guess 
values of  
$\langle n_{i,\uparrow} \rangle$ and $\langle n_{i,\downarrow} \rangle$. With 
the starting guess values of $\langle n_{i,\sigma} \rangle$ we diagonalize 
the Hamiltonians $H_{c,\uparrow}$ and $H_{c,\downarrow}$, and compute a new 
set of 
values for $\langle n_{i,\sigma} \rangle$. Next we replace the initial guess
values with the new set of values of $\langle n_{i,\sigma} \rangle$ and repeat
the process until the values of all $\langle n_{i,\sigma} \rangle$ converges. 
Substituting the final set of values in the Hamiltonians we calculate the 
two-terminal transmission probability using the Landauer formula. 
Transmission probability for up or down spin evaluated separately in terms of
Green's function from the relation~\cite{datta1}
\begin{equation}
T_{\sigma}(E)=Tr[\Gamma_S^{\sigma} G_{chain,\sigma}^r \Gamma_D^{\sigma} 
G_{chain,\sigma}^a]   
\label{eq7}
\end{equation}
where $\Gamma_{S(D)}^{\sigma}$ is the coupling matrix bearing the imaginary 
part of the self-energy $\Sigma_{S(D)}^{\sigma}$, arising due to the coupling 
between chain and the semi-infinite leads.
$G_{chain,\sigma}^r$ and $G_{chain,\sigma}^a$ are the retarded and advanced
Green's functions of the chain which include the effect of electrodes.
Thus we can write~\cite{datta1} 
$G_{chain,\sigma}^r=(E-H_{c,\sigma}-\Sigma_S^{\sigma}-
\Sigma_D^{\sigma})^{-1}$. For complete derivation of self-energies and 
effective Green's function ($G_{chain,\sigma}$) of the chain, see 
Appendix~\ref{self} and Ref.~\cite{datta1}.
Determining transmission probabilities of up and
down spin electrons we can calculate the total transmission probability 
from the relation $T(E)=\sum \limits_{\sigma} T_{\sigma}(E)$. Finally,
we evaluate the average density of states (ADOS) using the relation 
$\rho(E)=-(1/N \pi) Tr[Im[G_{chain}^r]]$. 

During the computation we use $\lambda=1$ in unit of 
$t$ (eV), except for Fig.~\ref{f5} where $\lambda$ is varying, and 
set $Q=(1+\sqrt 5)/2$. The other common parameters are as follows: 
$\tau_S=\tau_D=1\;$eV, $t=1\;$eV, $\epsilon_0=0$ and $t_0=2\;$eV. 
The on-site energy ($\epsilon_{i,\uparrow}^{\prime\prime}$ or
$\epsilon_{i,\downarrow}^{\prime\prime}$) in the chain, a sum of tree 
terms ($\lambda \cos(Qia)$, $\epsilon_i(V)$ and 
$U\langle n_{i,\downarrow}\rangle$ or $U\langle n_{i,\uparrow}\rangle$),
on the other hand is not a constant, and therefore, we cannot assign a
common value of it.  Throughout the calculations we measure 
all energies in unit of $t$ and bias voltage in unit of Volt ($V$) and 
set $c=e=h=1$, for simplification.

\section{Numerical Results and Discussion}

In this section we present the results based on the above theoretical 
formulation. 
\begin{figure}[ht]
{\centering \resizebox*{7cm}{6cm}{\includegraphics{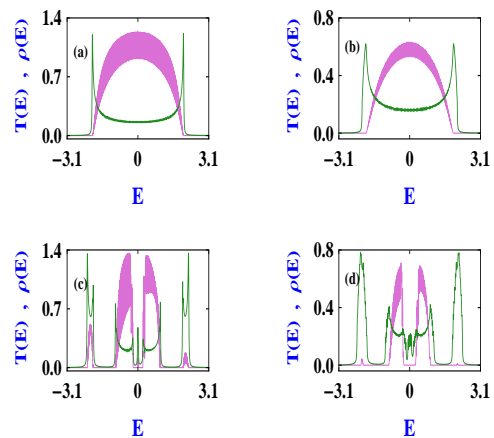}}\par}
\caption{(Color online). Total transmission probability (pink color)
and average DOS (green color) vs. energy $E$ for a non-interacting
chain considering $N=200$ and $U=0$, where (a) and (b) correspond to 
the chain with identical site potentials, while (c) and (d) represent 
the Aubry chain. For the left column we set $V=0$, and, it is $0.2$ 
for the right column. The results are computed for the linear bias 
drop across the chain.}
\label{f3}
\end{figure}
First we study the case of electron transmission through a non-interacting 
AA chain in presence of bias voltage and then we consider effect of 
Hubbard interaction into it.

Figures~\ref{f3}(a) and (b) present the results for the case of a 
non-interacting 1D chain in absence of incommensurate potential i.e., the
chain becomes a perfect $1$D lattice. For such a case we choose the bare
site potentials to zero, without loss of generality. 
In Fig.~\ref{f3}(a) we set the bias voltage $V$ to zero, while it is 
fixed at $0.2$ in Fig.~\ref{f3}(b).
In absence of external bias voltage all the energy eigenstates are extended 
in nature, and therefore, the transmission probability is finite at all 
\begin{figure}[ht]
{\centering \resizebox*{7cm}{6cm}{\includegraphics{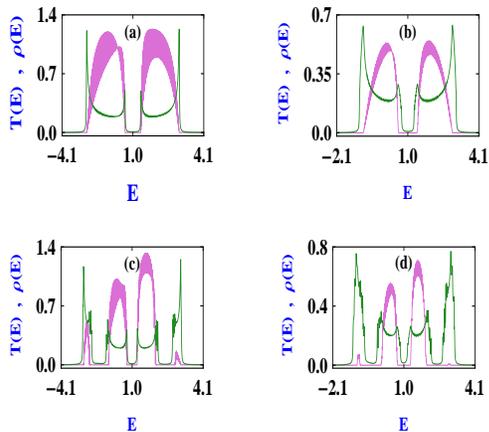}}\par}
\caption{(Color online). Total transmission probability and average DOS 
as a function of energy $E$ for an interacting chain with $U=2$, where 
(a)-(d) correspond to the identical meaning as in Fig.~\ref{f3}. All
the other parameters are same as considered in Fig.~\ref{f3}.}
\label{f4}
\end{figure}
\begin{figure}[ht]
{\centering \resizebox*{7cm}{6cm}{\includegraphics{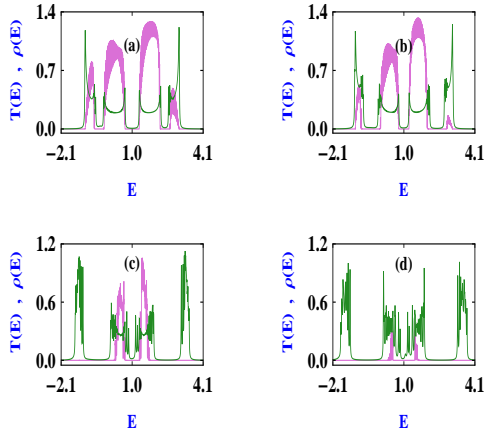}}\par}
\caption{(Color online). $T$-$E$ and $\rho(E)$-$E$ curves for a $200$ 
site interacting chain ($U=2$) for different values of $\lambda$, where
(a), (b), (c) and (d) correspond to $\lambda=0.5$, $1$, $2$ and $2.5$,
respectively. For all these cases we set $V=0$.}
\label{f5}
\end{figure}
energy eigenvalues. In presence of finite bias eigenstates at the band edges 
are no longer extended as it is evident from Fig.~\ref{f3}(b) that the 
transmission probability is zero while inside the band eigenstates are 
extended as we have finite transmission probability. It indicates that the 
choice of Fermi energy is quite important. If it is chosen to lie well 
inside the energy band the chain will be of conducting nature, while if it
lies near the band edges the chain behaves as an insulator. Such sharp 
transitions of the conducting behavior gives an idea of existence of 
mobility edges in presence of a finite bias. In the same figure, cases (c) 
and (d) represent the transmission probability and ADOS of a non-interacting 
AA chain in absence and presence of a linear bias drop. Presence of 
incommensurate potential leads to splitting of band. We see that there are 
two gaps embedded inside three bands. In presence of bias voltage, 
say $V=0.2$, the energy eigenstates belonging to the outermost two bands 
have negligible contribution to transmission unlike those of the middle band, 
and hence extended energy eigenstates as well as the localized ones are 
present leading to metal-insulator transition. The more we increase the 
bias voltage, more number of localized states will appear and at one 
stage all the states will be localized. 

Next we study the interplay between electric field and the Aubry ordering 
in 1D Hubbard chain. The transmission characteristics together with ADOS
\begin{figure}[ht]
{\centering \resizebox*{6.0cm}{3.5cm}{\includegraphics{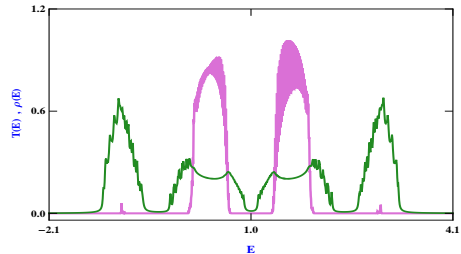}}\par}
\caption{(Color online). $T$-$E$ and $\rho(E)$-$E$ characteristics for 
an interacting AA chain considering $300$ atomic sites with $U=2$ and 
$\lambda=1$. The results are shown for a linear bias drop when $V=0.3$.}
\label{flin}
\end{figure}
\begin{figure}[ht]
{\centering \resizebox*{7cm}{7cm}{\includegraphics{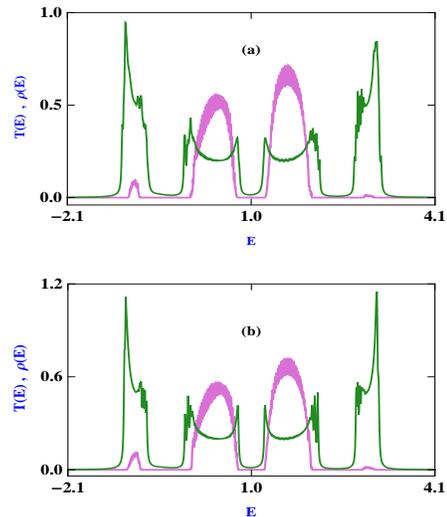}}\par}
\caption{(Color online). $T$-$E$ and $\rho(E)$-$E$ characteristics for 
an interacting AA chain with $N=200$, $U=2$ and $\lambda=1$ considering
non-linear bias drop across the chain. In (a) set the electrostatic potential
profile as given by the pink line in Fig.~\ref{efl}, while in (b) we choose
it according to the blue line of Fig.~\ref{efl}. Here we fix $V=0.2$.}
\label{f6}
\end{figure}
of an AA interacting chain are shown in Fig.~\ref{f4}. First row represents 
$T(E)$ and $\rho(E)$ as functions of $E$ for an interacting chain in absence 
of Aubry 
potential. In the half-filled case we have a Mott insulator (Fig.~\ref{f4}(a)) 
with a single gap at the band centre. For the $1$D AA Hubbard chain, we 
have a Mott gap at the center of the band, but additional gaps appear due 
to Aubry potential as clearly seen from Fig.~\ref{f4}(c). In absence of 
bias voltage, all the eigenstates of the periodic as well as the AA 
Hubbard chains have finite electron transmission probability. On the other
hand, in presence of finite bias localized states appear at the band edges 
both in case of a $1$D periodic Hubbard chain (Fig.~\ref{f4}(b)) and $1$D AA 
Hubbard chain (Fig.~\ref{f4}(d)). 

To investigate the precise role of $\lambda$ on transmission and ADOS we 
present results for different values of $\lambda$ in Fig.~\ref{f5} setting
$V=0$. We see that for $\lambda < 2t$, all the eigenstates are of extended 
in nature even in presence of $U$. The fact that for $\lambda < 2t$ all 
the eigenstates of $1$D AA chain behave like extended states and their 
behavior remains unaltered even for 
$U \neq 0$ which can be noticed from Figs.~\ref{f5}(a) and (b).
It is well known that for $\lambda = 2t$ states are critical and for
$\lambda > 2t$ all the eigenstates are localized, and for both these 
cases we have zero transmission probability. Quite interestingly from 
Figs.~\ref{f5}(c) and (d) we see that few conducting states appear in 
the middle of the inner two bands when $U \neq 0$. Physically it implies 
that electron-electron interaction changes the behavior of the AA chain.  

To test the invariant nature of the above discussed phenomena with 
respect to the parameter values, in Fig.~\ref{flin} 
we present the characteristics features of transmission probability 
together with average density of states considering a chain with 
different set of parameter values where $N=300$ and $V=0.3$. From the
spectra it is clear that all the physical pictures remain
unchanged and certainly it strengthens the invariant character of 
our analysis and can be verified experimentally.

Till now we have shown all the cases in presence of linear voltage drop 
across the chain. To get an idea regarding the behavior of transmission and 
average density of states in presence of non-linear bias drop let us focus
on the results given in Fig.~\ref{f6}. Two different non-linear profiles 
are taken into account following the curves (pink and blue lines) shown
in Fig.~\ref{efl}. For these two cases we also get similar kind of band
splitting and localization phenomenon, but a careful observation suggests 
that the transmission probability becomes higher for the flatter profile
(blue line of Fig.~\ref{efl}) compared to the other (pink line of 
Fig.~\ref{efl}) one. With increasing the flatness the localization effect 
due to electric field decreases, and, for the limiting case i.e., when the
bias drop takes place only at the two edges of the chain, transmission 
probability will be maximum when all the other parameters are kept 
unchanged.

\section{Conclusion}   

In the present work we critically investigate the role of electric field,
developed due to external bias, in an interacting $1$D Aubry chain. The 
interaction parameter is described within a Hartree-Fock mean field level
under tight-binding framework where transmission probability and ADOS are
evaluated from Green's function approach. The interplay between Aubry 
lattice, Coulomb correlation and electric field provides multiple mobility 
edges at different energies. Under this situation if we scan throughout the 
energy band window then electrons can allow to pass from source to drain
via the selective conducting energy channels providing finite electron
transmission, while for all other cases we get the insulating phase since 
then no electron can transmit through the localized channels. This phenomenon
clearly emphasizes that the present model can be utilized as a selective 
switching device.

\appendix
\section{Evaluation of self-energies and effective Green's function of 
the chain coupled to source and drain electrodes}
\label{self}

Since there is no spin flip mechanism (viz, from up spin to down spin 
or vice versa) in our problem, the net transmission probability is 
obtained from the relation $T(E)=\sum_{\sigma} T_{\sigma}(E)=
T_{\uparrow} + T_{\downarrow}$. To find $T_{\sigma}$ we need to evaluate
$H_{c,\sigma}$, $\Sigma_S^{\sigma}$, $\Sigma_D^{\sigma}$ and the
effective Green's function $G_{chain,\sigma}$. The Hamiltonian 
$H_{c,\sigma}$ is determined from the mean field scheme which is 
clearly described in Sec. II, and therefore, here we discuss elaborately
how to calculate other factors i.e., self energies and effective Green's 
function. For this, we can now ignore the spin index, as a matter of 
simplification, because the prescription is same for both the two spin 
cases.

Following the definition we can write the Green's function of the full system,
\begin{equation}
G^r = \left[(E + i\eta)I - H \right]^{-1}.
\label{gf42}
\end{equation}
where, $I$ is the identity matrix and $\eta \rightarrow 0^+$.
But there is a problem with Eq.~\ref{gf42}. Here we are working with an
open system i.e., a conductor connected with two semi-infinite electrodes. 
Therefore, $H$ has infinite dimension and so also $G^r$. Now it is not
possible to do any calculation with a matrix whose dimension is infinity.
So, to get rid of this situation we apply the partitioning technique
which maps the Green's function matrix in the reduced Hilbert space
of the conductor itself, and the effects of the side attached leads are
incorporated there.

Suppose we are considering a conductor attached to 
electrode $p$. Hence, the total Hamiltonian of the system can be written 
in a matrix form as,
\begin{equation}
H =
\left( \begin{array}{cc}
H_p & \tau_p \\
\tau_p^{\dagger} & H_c
\end{array}\right).
\label{gf43}
\end{equation}
Here, $H_c$ and $H_p$ are the Hamiltonian matrices describing the conductor 
and the side attached electrode. $\tau_p$ corresponds to the coupling 
matrix due to coupling of the conductor to the side attached electrode.

Hence, the Green's function is,
\begin{eqnarray}
G^r & = & \left( (E + i \eta)I - H \right)^{-1}\nonumber\\
& = & \left( \begin{array}{cc}
(E + i \eta)I - H_p & -
\tau_{p}\\
-\tau_{p}^{\dagger} & E I-H_{pc}
\end{array} \right)^{-1}
\label{gf44}
\end{eqnarray}
Now, partitioning the Green's function matrix in the same way like
the Hamiltonian matrix, we can rewrite the above equation as,
\begin{eqnarray}
\left(\begin{array}{cc}
G_p & G_{pc}\\
G_{cp} & G_c
\end{array} \right) 
&=&
\left( \begin{array}{cc}
(E + i \eta)I - H_p & -
\tau_{p}\\
-\tau_{p}^{\dagger} & E I-H_c
\end{array} \right)^{-1}\nonumber
\end{eqnarray}
i.e.,\begin{eqnarray}
\left(\begin{array}{cc}
G_p & G_{pc}\\
G_{cp} & G_c
\end{array} \right) 
\left( \begin{array}{cc}
(E + i \eta)I - H_p & -
\tau_{p}\\
-\tau_{p}^{\dagger} & E I-H_c
\end{array} \right)
&=&
\left( \begin{array}{cc}
1 & 0\\
0 & 1
\end{array} \right) \nonumber \\
\end{eqnarray}
We obtain two decoupled equations from the above equation, which are as
follows,
\begin{equation}
G_{cp}\left[ (E + i \eta)I - H_p\right] - G_c\tau_p^{\dagger} = 0
\label{gf46}
\end{equation}
and,
\begin{equation}
-G_{cp} \tau_p + G_c(E I -H_c) = 1
\label{gf47}
\end{equation}
We define, $(E + i \eta)I - H_p = (g_p^r)^{-1}$. So, from Eq.~(\ref{gf46}) 
we get,
\begin{eqnarray}
G_{cp}(g_p^r)^{-1} & = & G_c \tau_p^{\dagger} \nonumber\\
\Rightarrow G_{cp} & = & G_c\tau_p^{\dagger} g_p^r
\label{gf48}
\end{eqnarray}
Using the expression of $\mbox{\boldmath$G_{cp}$}$ from Eq.~(\ref{gf48})
in Eq.~(\ref{gf47}), we get,
\begin{eqnarray}
& & -G_c \tau_p^{\dagger} g^r_p \tau_p + G_c (E I - H_c) = 1\nonumber\\
& \Rightarrow & G_c \left[ E I - H_c - \tau_p^{\dagger} g^r_p 
\tau_p \right] = 1 \nonumber\\
& \Rightarrow & G_c \left[ E I - H_c - \Sigma_p \right] = 1 \nonumber\\
& \Rightarrow & G_c = \left[ E I - H_c - \Sigma_p \right]^{-1} 
\label{gf49}
\end{eqnarray}
where,
\begin{equation}
\Sigma_p = \tau_p^{\dagger} g^r_p \tau_p
\label{gf50}
\end{equation}
Here, $G_c (=\mathcal{G})$ is the effective Green's function which 
incorporates the effect of the electrode attached with the conductor.

In Eq.~\ref{gf49}, all matrices have the same dimension ($C$ $\times$ $C$),
where, $C$ is the dimension of the conductor. But at first sight it might
appear that the problem of inverting an infinite dimensional matrix still
remains, while evaluating $g^r_p$ to obtain the expression for $\Sigma_p$. 
But fortunately for an isolated infinite lead $g^r_p$ can be calculated 
analytically.

Now for $ij$-th element of the self-energy matrix,
\begin{eqnarray} 
\Sigma_{ij} & = & \langle i | \tau_p^{\dagger} g_p^r \tau_p|j \rangle 
\nonumber\\
& = & \sum_{m,n} \langle i | \tau_p^{\dagger} | m \rangle
\langle m | g_p^r | n \rangle \langle n | \tau_p| j \rangle \nonumber\\
& = & t^2 g_p^r (p_i,p_j).
\label{gf51}
\end{eqnarray}
Here, $g_p^r(p_i,p_j)$ denotes the ($p_i,p_j$)-th element of the matrix 
$g_p^r$. To work with Eq.~\ref{gf49} we have to reconstruct the $\Sigma_p$ 
matrix in $C$ $\times$ $C$ dimension. Here, all elements of $\Sigma_p$ 
matrix would be zero, except at the points ($i$,$j$) inside the conductor, 
which are adjacent to points ($p_i$,$p_j$) inside the electrode.

For more than one electrodes we have to add the effects of individual 
electrodes. It means if we have $p$ number of side-attached electrodes, 
then the effective Green's function will be,
\begin{equation}
\mathcal{G} = [E I - H_c - \sum \limits_p \Sigma_p]^{-1}
\label{gf52}
\end{equation}
From this expression we can easily write the desired effective Green's 
function for our two-terminal system as 
$\mathcal{G}=G_{chain}=(E-H_c-\Sigma_S-\Sigma_D)^{-1}$ which includes
the effects of both source and drain electrodes. For two different spin
sub-spaces it can be generalized as
$G_{chain,\sigma}=(E-H_{c,\sigma}-\Sigma_S^{\sigma}-\Sigma_D^{\sigma})^{-1}$.

\end{document}